\documentclass{aa}
\usepackage{graphicx}
\usepackage{txfonts}
\usepackage{ulem}
\begin{document}
\title{
Imaging the oxygen-rich disk toward the silicate carbon star EU~And
\thanks{
This work is based [in part] on observations made with the Spitzer 
Space Telescope, which is operated by the Jet Propulsion Laboratory, 
California Institute of Technology under a contract with NASA.
}
}

\author{Keiichi~Ohnaka\inst{1} 
\and
David~A.~Boboltz\inst{2} 
}

\offprints{K.~Ohnaka}

\institute{
Max-Planck-Institut f\"{u}r Radioastronomie, 
Auf dem H\"{u}gel 69, 53121 Bonn, Germany\\
\email{kohnaka@mpifr-bonn.mpg.de}
\and
United States Naval Observatory, 3450 Massachusetts Avenue, NW, 
Washington, DC 20392-5420, U.S.A.\\
\email{dboboltz@usno.navy.mil}
}

\date{Received / Accepted }

\abstract
{Silicate carbon stars are characterized by oxygen-rich circumstellar 
environments as revealed by prominent silicate emission, despite their 
carbon-rich photospheres.  
While the presence of a circumbinary disk or a disk around an unseen, 
low-luminosity companion is suggested to explain the peculiar dust 
chemistry, the origin of silicate carbon stars is still a puzzle to date.  
}
{We present multi-epoch high-angular resolution 
observations of 22~GHz \mbox{H$_2$O}\ masers toward the silicate carbon 
star EU~And to probe the spatio-kinematic distribution of oxygen-rich 
material. 
}
{EU~And was observed at three epochs (maximum time interval of 
14~months) with the Very Long Baseline Array (VLBA).  
}
{
Our VLBA observations of the 22~GHz \mbox{H$_2$O}\ masers have revealed 
that the maser spots are distributed along a straight line across 
$\sim$20~mas, with a slight hint of an S-shaped structure.  The observed 
spectra show three prominent velocity components at \mbox{$V_{\rm LSR}$}\ 
= $-42$, $-38$, and $-34$~\mbox{km s$^{-1}$}, with the masers in SW 
redshifted and those in NE blueshifted.  The maser spots located 
in the middle of the overall distribution correspond to the 
component at \mbox{$V_{\rm LSR}$}\ = $-38$~\mbox{km s$^{-1}$}, which 
approximately coincides with the systemic velocity.  
These observations can be interpreted as either an emerging helical 
jet or a disk viewed almost edge-on (a circumbinary or circum-companion 
disk). 
However, the outward motion measured in the VLBA images taken 14 months 
apart is much smaller than that expected from the jet scenario. 
Furthermore, the mid-infrared spectrum obtained with the Spitzer Space 
Telescope indicates that the 10~\mbox{$\mu$m}\ silicate emission is 
optically thin and the silicate grains are of sub-micron size.  
This lends support to the presence of a circum-companion disk, 
because an optically thin circumbinary disk consisting of such small 
grains would be blown away by the intense radiation pressure of the 
primary (carbon-rich) star.  
If we assume Keplerian rotation for the circum-companion disk, 
the mass of the companion is estimated to be 0.5--0.8~\mbox{$M_{\sun}$}.  
We also identify \mbox{CO$_2$}\ emission features at 13--16~\mbox{$\mu$m}\ 
in the Spitzer spectrum of EU~And---the first unambiguous detection of 
\mbox{CO$_2$}\ in silicate carbon stars.  
}
{}


\titlerunning{VLBA observations of EU~And}
\authorrunning{Ohnaka et al.}
\maketitle

\section{Introduction}
\label{sect_intro}

Oxygen-rich circumstellar material (i.e., silicate and/or Al$_2$O$_3$) is 
usually associated with M-type asymptotic giant branch (AGB) stars, 
reflecting their photospheric chemical composition.  
Surprisingly, however, carbon stars showing silicate emission (so-called 
``silicate carbon stars'') were discovered by IRAS (Little-Marenin 
\cite{little-marenin86}; Willems \& de Jong \cite{willems86}).  
The subsequent detection of H$_2$O and OH masers toward 
some of the silicate carbon stars, if not all, confirmed 
the oxygen-rich nature of the circumstellar material 
(e.g., Nakada et al. \cite{nakada87}, \cite{nakada88}; 
Benson \& Little-Marenin \cite{little-marenin87}; 
Little-Marenin et al. \cite{little-marenin88}; 
Barnbaum et al. \cite{barnbaum91}; Engels \cite{engels94}).  
On the other hand, optical spectroscopic studies show 
that $^{12}$C/$^{13}$C ratios in silicate carbon stars are as low as 
4--5 and thus they are classified as ``J-type'' carbon stars 
(e.g., Ohnaka \& Tsuji \cite{ohnaka99}), which is difficult to explain 
by the standard stellar evolution theory.

One currently accepted hypothesis 
suggests that silicate carbon stars have a low-luminosity companion 
(possibly a main-sequence star or a white dwarf) 
and that oxygen-rich material was shed by mass loss when the 
primary star was an oxygen-rich giant.  The oxygen-rich material 
is stored in a circumbinary disk even after 
the primary star becomes 
a carbon star (Morris \cite{morris87}; Lloyd-Evans \cite{lloyd-evans90}).  
Alternatively, Yamamura et al. (\cite{yamamura00}) propose that 
the oxygen-rich 
material is stored in a circumstellar disk around the companion. 
In this scenario, the observed silicate emission cannot originate directly
from the circum-companion disk heated by the companion itself, because
its luminosity is too low.  Instead,  Yamamura et al. (\cite{yamamura00}) 
argue that radiation pressure from the primary, carbon-rich, AGB star 
drives and heats an outflow from the circum-companion disk.  The 
observed silicate emission originates from this outflow.

Our recent $N$-band (8--13~\mbox{$\mu$m}) spectro-interferometric 
observations of the silicate carbon star IRAS08002-3803 using the ESO's 
Very Large Telescope Interferometer (VLTI) have spatially 
resolved the dusty environment of a silicate carbon star 
for the first time, and our radiative 
transfer modeling shows that the $N$-band visibilities 
can be fairly explained by an optically thick circumbinary disk in which 
small ($\sim$0.1~\mbox{$\mu$m}), amorphous silicate and a second grain 
species---amorphous carbon, large silicate grains ($\sim$5~\mbox{$\mu$m}), 
or metallic iron---coexist (Ohnaka et al. \cite{ohnaka06}).  
The $N$-band visibilities observed on another silicate 
carbon star IRAS18006-3213 show a wavelength dependence 
very similar to IRAS08002-3803, 
suggesting that IRAS18006-3213 also has a circumbinary disk 
(Deroo et al. \cite{deroo07}).  

Some evidence of long-lived disks around 
silicate carbon stars has been discovered by radio observations as well.  
Jura \& Kahane (\cite{jura99}) detected narrow CO ($J$=2--1, 1--0) 
emission lines toward two silicate carbon stars EU~And and BM~Gem, which 
indicate the presence of a reservoir of orbiting gas.  
Szczerba et al. (\cite{szczerba06}) and Engels 
(priv. comm.) obtained high-resolution 22~GHz \mbox{H$_2$O}\ maser maps 
toward the silicate carbon star V778~Cyg using MERLIN and VLBA, 
respectively.  The \mbox{H$_2$O}\ maser distributions are linearly aligned 
with a slightly S-shaped structure, which 
Szczerba et al. (\cite{szczerba06}) and Babkovskaia et al. 
(\cite{babkovskaia06}) interpret as a warped circum-companion disk viewed 
almost edge-on.  Therefore, as discussed in Yamamura et al. (2000) and 
Ohnaka et al. (2006), there may be two classes of silicate carbon stars: 
systems with optically thick circumbinary disks and those with 
circum-companion disks.  

While direct detection of companions of silicate carbon stars is 
still difficult due to the huge luminosity contrast ($\sim \! \!
10^4$~\mbox{$L_{\sun}$}\ and $\la$1~\mbox{$L_{\sun}$}\ for the primary 
star and the companion, respectively), 
Izumiura (\cite{izumiura03}) and Izumiura et al. (\cite{izumiura07}) 
detected blue continuum emission at $\la$4000~\AA\ and Balmer lines 
showing P~Cygni profiles with an outflow velocity of 
$\ga$400~\mbox{km s$^{-1}$}\ toward BM~Gem, 
which strongly suggests the presence of an accretion disk around 
an unseen companion.  

Despite this observational progress, the formation mechanisms of the 
circumbinary or circum-companion disks as well as the peculiar 
photospheric chemical composition of silicate carbon stars are little 
understood.  
Radio interferometry observations of the maser emission produced by 
oxygen-bearing molecules provide us with an excellent opportunity to 
study the spatio-kinematic distribution of oxygen-rich gas 
around silicate carbon stars.  In this paper, we present the results of 
multi-epoch, high-resolution observations of the 
22~GHz \mbox{H$_2$O}\  masers toward the silicate carbon star EU~And.

\section{Observations}
\label{sect_obs}

\begin{figure*}[tbh]
\resizebox{\hsize}{!}{\rotatebox{-90}{\includegraphics{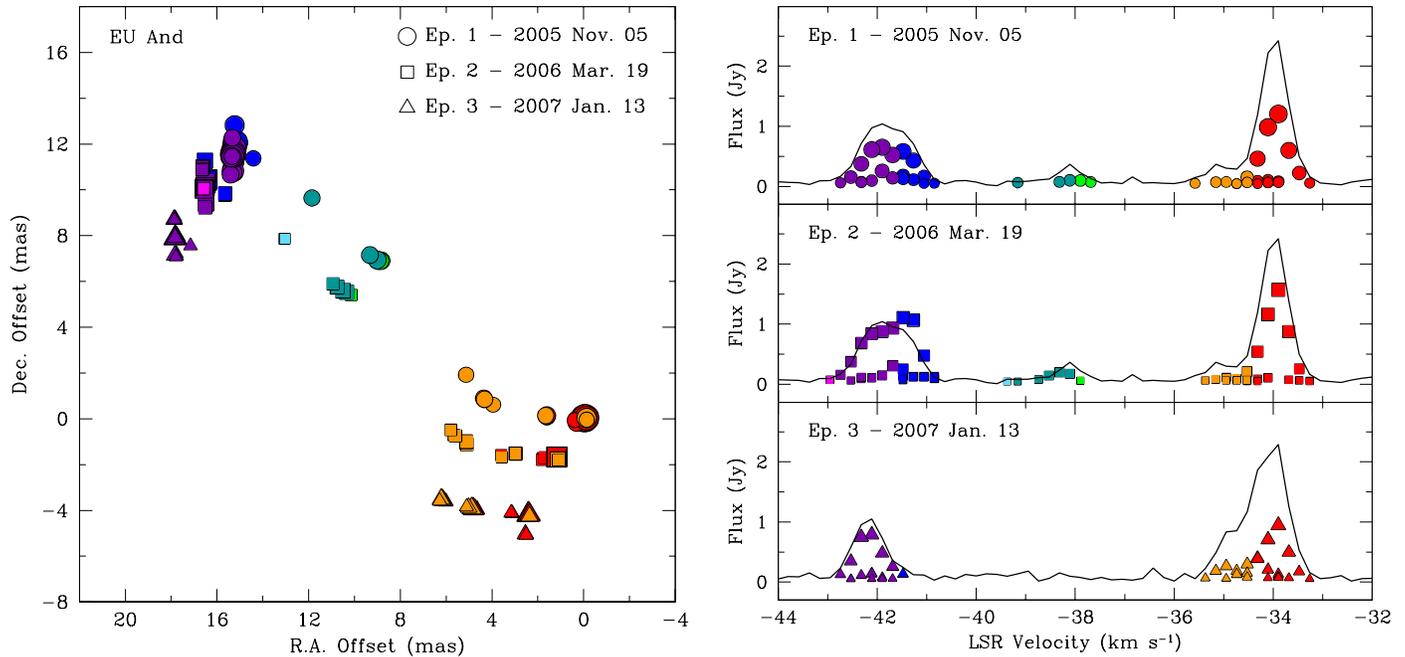}}}
\caption{
22~GHz \mbox{H$_2$O}\ masers observed toward EU~And with VLBA at three 
epochs.  The positions and velocities of the \mbox{H$_2$O}\ maser spots 
are derived by two-dimensional Gaussian fitting as described in 
Sect.~\ref{sect_obs}. 
{\bf Left:} 
Spatial distribution of the H$_2$O masers with point-color
representing the corresponding velocity bin in the spectrum and
point-size proportional to the logarithm of the maser flux density. 
The origin of R.A. and DEC. offsets is (23$^{h}$19$^{m}$58.8822$^{s}$, 
47\degr 14\arcmin 34.550\arcsec, J2000.0).  
The maser spots observed at three epochs are positionally offset 
because of proper motion and parallax.  
{\bf Right:}
Spectra formed by plotting maser component flux density 
versus LSR velocity, color-coded according to maser velocity.
The solid lines represent vector-averaged cross-power spectra 
on the Los Alamos--Pie Town VLBA baseline.  
A color version of this figure is available in the electronic 
edition.  
}
\label{obs_map}
\end{figure*}

The \mbox{H$_2$O}\ maser emission associated with EU~And 
($\alpha$=23$^h$19$^{m}$58.8814$^{s}$, 
$\delta$=47\degr 14\arcmin 34.567\arcsec, J2000.0, 
Epoch 2000.0, 
NOMAD Catalog, Zacharias et al. \cite{zacharias03})\footnote{This 
is offset by 32\arcsec\ from the position used for the VLA observation on 
1990 June 2 by Colomer et al. (\cite{colomer00}), who detected no 
\mbox{H$_2$O}\ 
maser toward EU~And.  The negative detection may have been due to this 
positional offset, although it is still possible that the \mbox{H$_2$O}\ 
masers were indeed absent in June 1990.  
} 
was observed at a rest frequency of 22.23508~GHz, using the 
Very Long Baseline Array (VLBA) operated 
by the National Radio Astronomy Observatory (NRAO)\footnote{The 
National Radio Astronomy Observatory is a facility of the National 
Science Foundation operated under cooperative agreement 
by Associated Universities, Inc.}.  
The observations were carried out over three epochs 
spanning approximately 14~months: 
2005 November 5, 2006 March 19, and 2007 January 13.  
VLBA's 10 antennas were used at the first and second epochs, 
while only 9 antennas could be used at the third epoch due to a technical 
problem of the St. Croix station.
For each 5-hr epoch, EU~And and two extragalactic calibrator sources
(3C454.3 and J2322+5057) were observed.  We used the technique 
of rapid switching between the target source and the nearby phase 
reference source, J2322+5057, in order to remove residual phase 
offsets in the target source due to the atmosphere.  
The data were recorded in dual circular polarization with the 8-MHz band 
centered on a local standard of rest (LSR) velocity of 
$-36.0$~\mbox{km s$^{-1}$}.  
The system temperatures and the sensitivities were of the order of 
100~K and 10~Jy~K$^{-1}$, respectively, for all three epochs.  

The data were correlated at the VLBA correlator in Socorro, 
New Mexico, which produced auto- and cross-correlation spectra 
with 512 channels, corresponding to a channel spacing of 15.63~kHz 
(0.22~\mbox{km s$^{-1}$}).  For the calibration of the correlated data, 
we followed the standard reduction procedures for VLBA spectral line 
experiments using the Astronomical Image Processing System (AIPS) 
maintained by NRAO.  
A bandpass calibration was carried out using intermittent 
(every 30 minutes) scans on 3C453.3.  A fringe fit was performed 
on the phase-reference source in order to remove effects due to 
instrumentation and atmosphere on the phase not removed 
in the correlation.  The resulting residual phase delays, phase rates, and 
phases were applied to the target source.  At this point a preliminary
image cube was produced for EU And.  Because the signal-to-noise
in these images was not optimal, we performed a second fringe fit
on a strong reference maser feature 
(at \mbox{$V_{\rm LSR}$}\ = $-34$~\mbox{km s$^{-1}$}\ ) 
in the spectrum of EU~And and applied the residual phase rates
to the target.    Finally, an iterative self-calibration and imaging 
procedure was performed on the reference channel and the solutions for the 
residual phase and amplitude corrections were applied to all spectral 
channels.  Images with 1024$\times$1024 pixels (61$\times$61~mas) 
were produced for all velocity channels between \mbox{$V_{\rm LSR}$}\ 
= $-24$ and $-52$~\mbox{km s$^{-1}$}\ using beam sizes of 
0.71$\times$0.29~mas, 0.73$\times$0.36~mas, and 0.82$\times$0.31~mas for 
the first, second, and third epochs, respectively.  
Typical rms off-source noise in the final images is 7--10~mJy~beam$^{-1}$.

Maser components were identified, and their positions were measured 
by fitting the emission in the final images of each spectral channel with 
two-dimensional (2-D) Gaussians using the AIPS task SAD. 
The errors of the positions (within an epoch) 
are 10--20~$\mu$as for the strong maser features 
and approach $\sim$50~$\mu$as for the weaker features.  
The absolute astrometric information obtained by phase-referencing 
is lost in the above self-calibration procedure.  
To restore this information to the final component positions, we fit 2-D 
Gaussians to the strongest maser feature (assumed to be the same for all 
three epochs at \mbox{$V_{\rm LSR}$}\ = $-34$~\mbox{km s$^{-1}$}) 
prior to any self-calibration.  The resulting shifts were applied to all 
maser components identified in the final images.  The absolute positions 
of the strongest maser obtained for the first, second, and third epochs 
are 
(23$^{h}$19$^{m}$58.8822$^{s}$, 47\degr 14\arcmin 34.550\arcsec), 
(23$^{h}$19$^{m}$58.8823$^{s}$, 47\degr 14\arcmin 34.548\arcsec), and 
(23$^{h}$19$^{m}$58.8824$^{s}$, 47\degr 14\arcmin 34.546\arcsec ), 
respectively (all in J2000.0).   

The accuracy of the absolute position of the strongest maser feature 
is primarily limited by errors resulting from the transfer of the residual 
phase corrections from the extragalactic reference source to the target 
source and by the error in the absolute position of the reference source 
itself.  In order to estimate the error involved in the transfer of the 
residual phase corrections, we essentially reversed the procedure used to 
determine the absolute position of the target maser source.  We applied 
the residual phase solutions resulting from the self-calibration on the 
target source (strongest maser at \mbox{$V_{\rm LSR}$}\ = 
$-34$~\mbox{km s$^{-1}$}\ ) to the extragalactic reference source.  
We then imaged the reference source and performed a 2-D Gaussian fit to 
determine its position.  The separation between the positions of the 
target/calibrator pair (i.e., the arc length) was computed for 1) both 
sources calibrated with the extragalactic reference source solutions, 
and 2) both sources calibrated with the target source solutions.  
The difference between the arc lengths yields an estimate of the error 
involved in the phase transfer.  The computation of the arc lengths is 
unnecessary in experiments specifically designed for astrometry where 
a second extragalactic check source is typically observed.  This
procedure was repeated for all three epochs and the standard 
deviation of the difference in arc lengths was found to be 0.12~mas.   
This error in the phase transfer compares favorably with the error in the 
absolute position of the reference source itself, which is a defining 
source of the International Celestial Reference Frame (ICRF) with an error 
of 0.26~mas (Fey et al. \cite{fey04}).  The total error in the absolute
position of the strongest maser feature is estimated from the 
root-sum-square of the two errors described above to be $\sim$0.3~mas.

\section{Results}
\label{sect_results}

Figure~\ref{obs_map} shows the spatio-kinematic 
distributions (left panel) and spectra 
(right panels) of the \mbox{H$_2$O}\ masers toward EU~And as observed 
over the three epochs.  
At all epochs, the \mbox{H$_2$O}\ maser spectra are characterized by two 
strong peaks at \mbox{$V_{\rm LSR}$}\ = $-34$~\mbox{km s$^{-1}$}\ and 
$-42$~\mbox{km s$^{-1}$}.  
In the first and second epochs, weak features were also detected 
at $-38$~\mbox{km s$^{-1}$}, approximately in the middle of the two peaks. 
The velocity of these weak maser features is in agreement with 
that of the narrow CO lines ($\sim \! -37$~\mbox{km s$^{-1}$}) measured by 
Jura \& Kahane (\cite{jura99}).  
The velocity of the weak maser features also coincides with the 
radial velocities derived from the optical spectra, which represent 
the velocity of the primary carbon-rich AGB star 
(Barnbaum \cite{barnbaum91}).  However, as discussed below, the putative 
companion, instead of the primary star, is more likely to be located at 
the position of these weak maser components.  
We note that while the spectra taken over 14~months appear rather 
stable, the \mbox{H$_2$O}\ masers toward EU~And exhibit remarkable 
temporal variations in strength and velocity.  
For example, Little-Marenin (\cite{little-marenin88}) 
detected maser components near 
\mbox{$V_{\rm LSR}$}\ = $-30$~\mbox{km s$^{-1}$}\ as strong as 8~Jy, 
which are non-existent in our VLBA data.

The \mbox{H$_2$O}\ masers at all three epochs are 
linearly aligned across $\sim$20~mas, with a slight hint of an S-shaped 
structure.  As explained in Sect.~\ref{sect_obs}, 
the offsets among the epochs shown in Fig.~\ref{obs_map}
are real and represent the proper motion and parallax. 
In fact, the slight curvature seen in the direction of the positional 
displacement over the three epochs, which is 
discernible particularly for the strong features (the one at $(0, 0)$ 
at the first epoch), shows the annual parallax component.  
The observed S-shaped structure can 
be interpreted as an edge-on disk or an emerging helical jet.  
Such a nearly linear distribution of the \mbox{H$_2$O}\ masers 
is similar to those observed toward the so-called water fountain 
sources (e.g., Imai et al. \cite{imai02}; Boboltz \& Marvel 
\cite{boboltz07} and references therein), which show well-collimated, 
fast jets of \mbox{H$_2$O}\ masers.  However, the outflow velocities of 
the \mbox{H$_2$O}\ masers toward the water fountain sources are much 
higher ($\sim$60--150~\mbox{km s$^{-1}$}) than that observed toward EU~And 
($\sim$5~\mbox{km s$^{-1}$}), although the projection effect is uncertain.

In order to examine the jet scenario, 
we measured the increase of the separation (arc length) between the 
masers associated with the two highest peaks of emission in the spectrum.  
Since a single maser 
spans multiple channels in the spectrum, we performed a flux density
squared weighted average over velocity and position prior to measuring
the separations between the two peak features.  The difference in the 
arc lengths between the first and third epochs is $\sim$0.3~mas. 
Using this measurement, we estimate the velocity $V$ 
and the inclination angle $i$ (angle toward us out of the plane of the 
sky) of a water-fountain-like outflow for EU~And as follows.  
The observed velocity separation of the two strongest masers is 
8~\mbox{km s$^{-1}$}, 
which means that $2 V \sin i = 8$~\mbox{km s$^{-1}$}.  
The spatial separation of the two 
masers increases by $2 V \cos i \times \Delta t /d = 0.3$~mas, where 
$\Delta t$ is the time interval between the first and third epochs 
(14~months), and $d$ is the distance of EU~And.  
With distances of 1.5--2.6~kpc adopted (Jura \& Kahane \cite{jura99}; 
Engels \cite{engels94}), we obtain $V$ = 4.1--4.3~\mbox{km s$^{-1}$}\ and 
$i$ = 68--77\degr.  
These outflow velocities are much lower than those observed toward 
the water-fountain sources, making the helical jet interpretation 
unfavorable, although the distance of EU~And may be remarkably larger 
than the above estimates and/or the outflow velocity of EU~And may 
indeed be so low due to some driving mechanism different from that in 
the water-fountain sources.  
A further long-term monitoring as 
well as a measurement of proper motion with VLBA would be necessary 
to settle this issue.

The \mbox{H$_2$O}\ maser distributions observed toward EU~And are similar 
to that observed toward V778~Cyg by Szczerba et al. (\cite{szczerba06}) 
and Engels (priv. comm.).  Our observations of 
another silicate carbon star IRAS07204-1032 using VLBA and 
the Very Large Array (VLA) have also revealed a similar spectrum 
and spatial distribution of the \mbox{H$_2$O}\ masers (Boboltz et al. in 
prep.).  
The expectation of observing linear distributions of masers
toward the three silicate carbon stars imaged to date may seem
statistically unlikely, given a random distribution of disk inclination
angles.  However, it is quite possible that this is due to a selection
effect in which masers are preferentially detected in objects
with the most favorable geometry for maser amplification (i.e., 
nearly edge-on disks).
A survey of more objects with VLA and VLBA would be useful for 
obtaining a definitive answer to the interpretation of the masers 
toward silicate carbon stars.

\section{Discussion}
\label{sect_discuss}

\begin{figure}
\resizebox{\hsize}{!}{\rotatebox{0}{\includegraphics{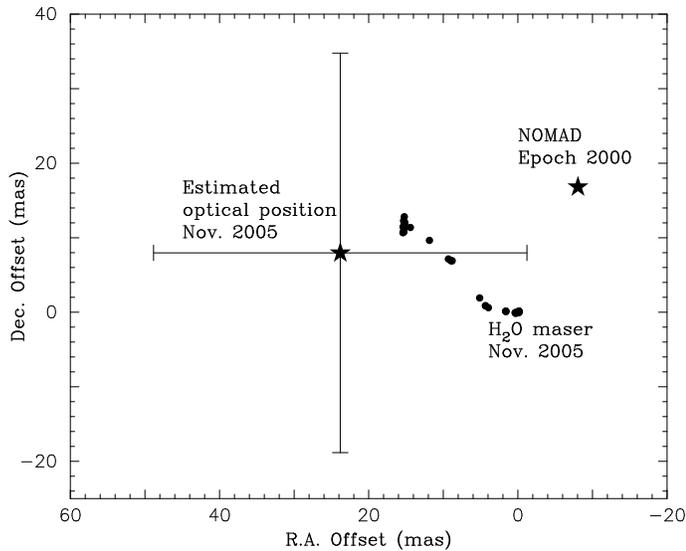}}}
\caption{
The optical position of EU~And in 2000 (NOMAD catalog) and the expected 
optical position at the first epoch of the VLBA observations (Nov. 2005).  
The error bars are the sum of the uncertainties of the absolute position 
and the proper motion (see Sect.~\ref{sect_discuss}).  
}
\label{euand_optpos}
\end{figure}

Although our \mbox{H$_2$O}\ maser maps of EU~And lend support to 
the presence of an edge-on disk, these observations alone cannot 
clarify whether the disk is located around the carbon-rich primary 
star (circum-primary disk) 
or the whole binary system (circumbinary disk) or an unseen 
companion (circum-companion disk).  
The optical position of EU~And, 
which corresponds to that of the primary star, is ($\alpha$, $\delta$) 
= (23$^{h}$19$^{m}$58.8814$^{s}$, +47\degr 14\arcmin 34.567\arcsec, 
J2000.0) with proper motion of ($\mu_{\alpha}$, 
$\mu_{\delta}$) = ($5.5 \pm 1.7$, $-1.5 \pm 2.0$) mas~yr$^{-1}$ 
(NOMAD catalog, Zacharias et al. \cite{zacharias03}).  
The position of the primary star expected for the first epoch, which  
is shown in Fig.~\ref{euand_optpos}, appears to be offset from the 
\mbox{H$_2$O}\ masers.  
However, the uncertainty of the absolute position 
($\pm 15$~mas, Zacharias et al. \cite{zacharias04}) as well as that of 
the proper motion is too large to examine whether the primary 
star is located at the center of the \mbox{H$_2$O}\ masers or indeed 
offset from the maser distribution.  

\subsection{Constraints from the mid-infrared spectrum}

\begin{figure}
\resizebox{\hsize}{!}{\rotatebox{0}{\includegraphics{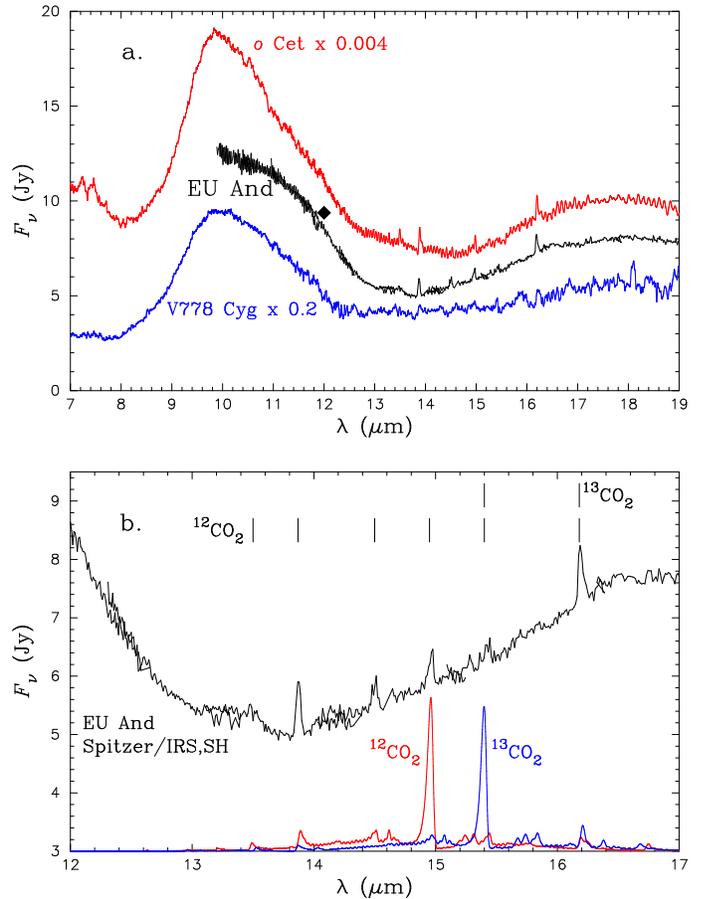}}}
\caption{
{\bf a:} 
Spitzer/IRS spectrum of EU~And, together with the scaled ISO/SWS spectra 
of V778~Cyg and $o$~Cet.  The diamond represents the IRAS 
12~\mbox{$\mu$m}\ flux of EU~And.  
{\bf b:} \mbox{CO$_2$}\ emission features identified in the Spitzer/IRS 
spectrum of EU~And are marked with the ticks.  
The scaled absorption cross section of $^{12}$CO$_2$ and $^{13}$CO$_2$ for 
1000~K are also shown.  
}
\label{euand_irs}
\end{figure}

On the other hand, as Yamamura et al. (\cite{yamamura00}) argue, 
a circum-primary or circumbinary disk would be unstable against the 
intense radiation pressure from the primary carbon-rich star 
($\sim$10$^3$--10$^4$~\mbox{$L_{\sun}$}), if the 
disk is entirely optically thin.  
In order to examine whether or not the silicate dust around EU~And is 
optically thin, we obtained its mid-infrared spectrum from the data 
archive of the Spitzer Space Telescope (Werner et al. \cite{werner04}).  
EU~And was observed on 
2004 December 9 (Program ID: P03235, P.I.: C.~Waelkens) 
with the InfraRed Spectrograph\footnote{
The IRS was a collaborative venture between Cornell University and Ball 
Aerospace Corporation funded by NASA through the Jet Propulsion Laboratory 
and Ames Research Center.
}
(IRS, Houck et al. \cite{houck04}) in the Short-High (SH) and Long-High 
(LH) modes with a spectral resolution of $\sim$600 .  
We downloaded the Basic Calibrated Data (BCD) processed with the 
S15-3 pipeline and extracted the spectrum using SMART v.6.2.5 
(Higdon et al. \cite{higdon04})\footnote{
SMART was developed by the IRS Team at Cornell University and is available 
through the Spitzer Science Center at Caltech.
}.  
Figure~\ref{euand_irs} shows the Spitzer/IRS spectrum of EU~And 
(no sky subtraction was performed) 
together with the spectra of the silicate carbon star V778~Cyg 
and the oxygen-rich AGB star $o$~Cet obtained with the Short 
Wavelength Spectrometer (SWS) onboard the Infrared Space Observatory 
(ISO).  As Yamamura et al. (\cite{yamamura00}) show, the 10~\mbox{$\mu$m}\ 
silicate features of the latter two stars indicate that the silicate 
emission is optically thin and silicate grains are of sub-micron 
size. 
Obviously, 
the spectrum of EU~And closely resembles those of V778~Cyg and $o$~Cet.  
A comparison with the IRAS 12~\mbox{$\mu$m}\ flux of EU~And and 
its Spitzer/IRS spectrum 
reveals that the flux level has little changed for the last 21 years, 
and EU~And is similar to 
V778~Cyg in this aspect as well.  
This means that the silicate emission of EU~And is optically 
thin with sub-micron grain sizes, 
rendering the possibility of a circum-primary or circumbinary disk 
unlikely.  Therefore, the disk-like structure discovered by the 
\mbox{H$_2$O}\ masers is likely to represent a circum-companion disk.

The predominance of sub-micron-sized grains appears to contradict the 
conclusion of Jura \& Kahane (\cite{jura99}) that the grain size around 
EU~And should be 
as large as $\ge$0.2~cm to be gravitationally bound against radiation 
pressure.  However, they assumed a circum-primary or circumbinary disk 
around the central star with a luminosity of $10^4$~\mbox{$L_{\sun}$}.  
In the case of a disk around a low-luminosity 
companion ($\sim$1~\mbox{$L_{\sun}$}), grains as small as 
0.2~\mbox{$\mu$m}\ are 
stable against the radiation pressure of the companion, 
although this does not entirely exclude the presence of larger grains 
due to grain growth in dense regions (e.g., mid-plane) of the 
circum-companion disk.

In the Spitzer/IRS spectrum of EU~And shown in Fig.~\ref{euand_irs}, 
we can identify the \mbox{CO$_2$}\ emission features.  
To illustrate this identification, 
we also show the (scaled) absorption cross sections of $^{12}$CO$_2$ and 
$^{13}$CO$_2$, which were calculated using the \mbox{CO$_2$}\ line list of 
the HITRAN database (Rothman et al. \cite{rothman05}) for 1000~K 
(the column densities of the both isotopic species are set to be equal).  
The \mbox{CO$_2$}\ features are clearly seen 
in the ISO/SWS spectrum of the oxygen-rich star $o$~Cet, and also 
marginally in V778~Cyg.  Yamamura et al. (\cite{yamamura00}) could 
only tentatively identify these \mbox{CO$_2$}\ features in V778~Cyg 
because of the low S/N ratio, and the case of EU~And is the first 
unambiguous detection of \mbox{CO$_2$}\ toward silicate carbon stars.  
This is another piece of evidence 
for oxygen-rich gas around silicate carbon stars, and the mid-infrared 
\mbox{CO$_2$}\ features will enable us to estimate the \mbox{CO$_2$}\ 
density and temperature if the disk structure is more tightly constrained 
in the future.  
It would also be possible to derive the $^{12}$C/$^{13}$C 
ratio of the circumstellar gas from an analysis of the mid-infrared 
$^{12}$CO$_2$ and $^{13}$CO$_2$ features, which would shed new light 
on the origin of the abnormally low $^{12}$C/$^{13}$C ratios in 
silicate carbon stars.  

It is worth noting that $o$~Cet (Mira~AB), whose mid-infrared spectrum 
closely resembles that of EU~And, is a well-known binary system 
(separation $\sim$0\farcs5) with a main-sequence or white dwarf 
companion.  
The high-resolution mid-infrared images of Mira~AB recently obtained 
by Ireland et al. (\cite{ireland07}) suggest that the edge of an accretion 
disk around Mira~B (companion) is heated by Mira~A 
(the primary star at the AGB).  
As Izumiura et al. (\cite{izumiura07}) suggest, 
this may resemble the circumstellar environment of some --- if not all --- 
silicate carbon stars (including EU~And).  
Unlike silicate carbon stars, the circum-companion disk around Mira~B 
consisting of silicate is embedded in the oxygen-rich outflow from Mira~A, 
which masks the mid-infrared spectroscopic signature of the 
circum-companion disk.  

\subsection{Nature of the companion}
\label{subsect_pv}

\begin{figure}
\resizebox{\hsize}{!}{\rotatebox{0}{\includegraphics{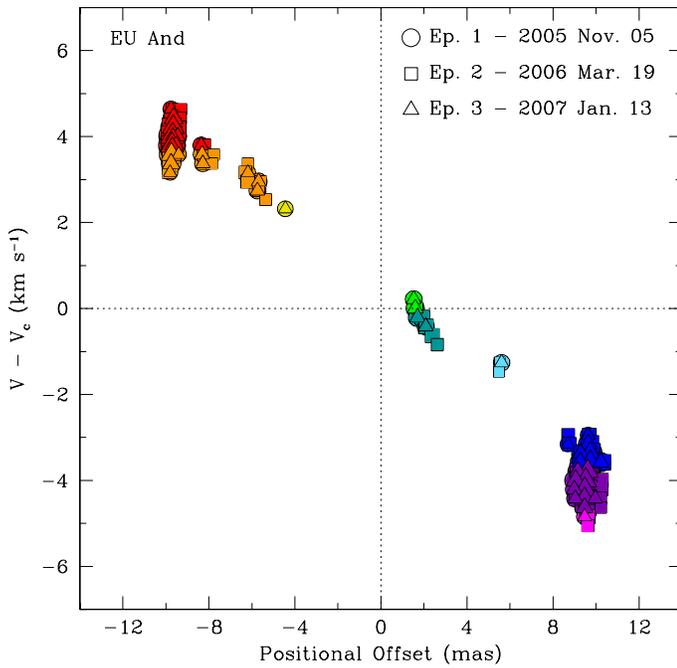}}}
\caption{
The $p$-$V$ diagram for the \mbox{H$_2$O}\ masers toward EU~And.  
The center position and velocity were derived as described in 
Sect.~\ref{subsect_pv}.  
The center position (for the first epoch) 
is located at $(7.7, 5.9)$~mas in the left panel 
of Fig.~\ref{obs_map}, while the center velocity $V_{c}$ is 
$-37.9$~\mbox{km s$^{-1}$}.  
The circles, squares, and triangles represent the data taken at the first, 
second, and third epochs, respectively, with the velocities represented 
by the colors as in Fig.~\ref{obs_map}.  
A color version of this figure is available in the electronic 
edition.  
}
\label{pv_diagram}
\end{figure}

In order to draw a $p$-$V$ diagram for the observed masers toward EU~And, 
we determined the center of the overall spatial distribution by taking the 
mid-point of a line connecting the mean positions of the redshifted and 
blueshifted masers at each epoch.  The center velocity was also derived 
by halving the difference between the mean velocities of the 
redshifted and blueshifted masers.  
The resulting $p$-$V$ diagram for EU~And is shown in 
Fig.~\ref{pv_diagram}.  It is very similar to that derived for 
V778~Cyg by Szczerba et al. (\cite{szczerba06}), although the number 
of maser spots is much smaller than for V778~Cyg.  
Both objects show a linear part with very large gradients at both ends of 
the velocity range.  This is consistent with the presence of a Keplerian 
disk, although it is not definitive evidence.  
If we assume that the (circum-companion) disk is in Keplerian 
rotation, and the center of the disk is located in the middle of the 
maser distribution at \mbox{$V_{\rm LSR}$}\ = $-37.9$~\mbox{km s$^{-1}$}, 
we can estimate the mass of the companion at the disk center.  
As shown in Pestalozzi et al. (\cite{pestalozzi04}), maser emission 
observed toward an edge-on disk has three maxima---on the line of sight 
toward the disk center and near the lines of sight tangential to 
the outer edge of the disk.  If the strong \mbox{H$_2$O}\ maser components 
to NE and SW correspond to the maxima near the outer edge of the disk, 
it follows that the Keplerian velocity is $\sim$5~\mbox{km s$^{-1}$}\ at a 
radius of 10~mas = 10 $\times$ ($d$/kpc) AU, where $d$ is the distance 
of EU~And.  This translates into 0.3 ($d$/kpc) \mbox{$M_{\sun}$}\ for the 
central mass (0.5--0.8~\mbox{$M_{\sun}$}\ for $d$ = 1.5--2.6~kpc).  
This mass would be reasonable for an unevolved main-sequence star 
or a white dwarf.  

It should be noted, however, that the strong maser components at $-42$ 
and $-34$~\mbox{km s$^{-1}$}\ may not represent the outer edge of the 
disk.  As mentioned above, the strong maser components at 
$\sim \! -30$~\mbox{km s$^{-1}$}\ 
were detected by Little-Marenin et al. (\cite{little-marenin88}), 
but the lack of spatial information does not allow us to take these 
observations into account in estimating the mass of the secondary 
star.  
Moreover, there may be masers fainter than the VLBA's detection limit 
and/or more diffuse emission which would be resolved out by VLBA.  
Therefore, the above estimate of the companion mass based on Keplerian 
rotation is a lower limit.  
The estimate of the companion mass also depends on the 
disk geometry assumed in the analysis.  For example, for the MERLIN data 
on V778~Cyg presented by Szczerba et al. (\cite{szczerba06}), 
Babkovskaia et al. (\cite{babkovskaia06}) applied a doubly warped disk 
model and derived a companion mass of 1.7~\mbox{$M_{\sun}$}, which is 
significantly larger than the 0.06~\mbox{$M_{\sun}$}\ derived by 
Szczerba et al. (\cite{szczerba06}).  This highlights, once again, 
the importance of long-term monitoring of the \mbox{H$_2$O}\ masers in 
order to derive the properties of the disk and the secondary star.

\section{Conclusion}
\label{sect_concl}

Our VLBA observations of the 22~GHz \mbox{H$_2$O}\ masers toward the 
silicate carbon star EU~And have revealed that the masers are linearly 
aligned with a slight hint of an S-shaped structure, with the masers in SW 
redshifted and those in NE blueshifted.  
Such a spatio-kinematic structure can be interpreted either as an edge-on 
disk or an emerging jet.  The \mbox{H$_2$O}\ maser maps obtained at three 
epochs over 14~months show a little outward motion of 0.3~mas, 
but the outflow velocity of $\sim$4~\mbox{km s$^{-1}$}\ 
estimated from this outward motion 
is too low compared to those observed toward the water-fountain sources.  
This lends support to the disk interpretation, 
although some kind of outflow (possibly with a velocity much lower 
than in the water-fountain sources) cannot be entirely ruled out due to 
the uncertainty in the distance.  
The mid-infrared spectrum of EU~And obtained with the Spitzer/IRS 
shows that the silicate emission is optically thin and emanates 
from sub-micron-sized grains, which suggests 
that the \mbox{H$_2$O}\ masers originate in a circum-companion disk seen 
nearly edge-on.  Furthermore, we unambiguously identified the 
\mbox{CO$_2$}\ features at 13--16~\mbox{$\mu$m}\ in the Spitzer spectrum 
for the first time.  

If we assume that the disk is in Keplerian motion, the mass of the 
putative secondary star is estimated to be 0.5--0.8~\mbox{$M_{\sun}$}.  
However, given the remarkable variability of the water masers 
and the possible presence of fainter or more diffuse emission 
not detected by VLBA, the estimated mass should be regarded as 
a lower limit.  Observations with a more compact, high-sensitivity,  
array (e.g., the VLA or MERLIN) might be useful in characterizing any such
extended emission.  Future VLBA observations will be necessary to 
put stronger constraints on the geometry of the disk and the properties of 
the secondary star.  Astrometric monitoring with the VLBA should also 
provide more accurate estimates of the parallax and proper motion of 
EU And, thus improving our understanding of the three-dimensional 
structure of the oxygen-rich gas surrounding the system.

\end{document}